# Uranium Critical Point Location Problem

Igor Iosilevskiy[a] and Victor Gryaznov[b]

[a] *Moscow Institute of Physics and Technology (State University) Russia*
[b] *Institute of Problem of Chemical Physics RAS, Chernogolovka, Russia*

**Abstract**
Significant uncertainty of uranium critical point parameters in present knowledge is considered. Paper is to reveal thermodynamic aspects of the problem through comparison of some available theoretical predictions for the uranium critical point parameters as well as to discuss in brief plausible versions to meet existing contradiction. New calculations of gas–liquid coexistence in uranium by modern thermodynamic code are included in the analysis.

## 1. Introduction

Parameters of critical point (CP) of metallic uranium and uranium-bearing compounds ($UO_{2\pm x}$, UC, UN, $UF_6$ *etc.*) are in urgent need, firstly, to develop perspective powerful devices (e.g., [1]), and, secondly, as an important ingredient for the analysis of nuclear safety in hypothetical severe reactor accidents at nuclear plants exploited recently (e.g., [2]). For most of metals including uranium both their critical temperature and pressure appear to be too high for precise experimental study (except heavy alkali and mercury). Thus, nowadays the CP parameters of uranium and uranium-bearing compounds are known mostly due to theoretical predictions. Various approaches have been used for this purpose (see reviews [3],[4] etc.). As a rule, all the approaches give rather close results for most of substances. That is why it looks tempting to consider the deviation in results of various estimations as a measure of uncertainty in knowledge of critical point parameters. Hence, each case of violation of this empirical tendency is valuable in view of reliable understanding of the critical point problem. This is just the case for uranium (and uranium dioxide) which represents remarkable exception from this empirical rule [5], i.e. outstanding *contradiction* between results of various approaches. Moreover, a similar contradiction proved to be valid [6] for the whole group of "bad" metals (Co, W, Mo etc.) with the precedent of uranium being the most prominent one. Despite of great applicative importance of uranium EOS, we are still not aware even approximately the parameters of high-temperature part of uranium gas-liquid coexistence including true parameters of its critical point. In search for the problem solving, it is essential (A) to disavow some results of one (or more) basic experiments on thermodynamic properties of liquid uranium or/and (B) to assume at least one (or more) significant anomaly in properties of gas-liquid phase transition in uranium. Present contribution is devoted to revealing of thermodynamic aspects of the problem, as well as to brief discussion of plausible variants for its possible resolutions.

## 2. EOS of uranium in applications

The high-temperature equation of state (EOS) of uranium, including its critical point, is of great importance for wide number of applications, in particular, design of non-traditional schemes for nuclear reactor. Such a perspective scheme has been developed recent decades [7],[1]. In contrast to a nuclear reactor with solid fuel exploited presently, the basic feature of so-called *gas-core nuclear reactor* (GCNR) is a high-temperature dense plasma state of uranium fuel at its work cycle ($T_U \sim 10^4 \div 10^5$ K, $p \sim 10 \div 10^2$ GPa). The principal advantage of GCNR is its ability to heat working fluid up to the considerable temperature level of $T_{WF} \sim 10^3 \div 10^4$ K. Such working fluid could be effectively used afterwards in rocket engine or in MHD energy converters, etc. [8].

High-temperature uranium EOS is also required for investigation of a nuclear safety problem to construct the global uranium-oxygen phase diagram [9],[10],[12],[13] (see also [11],[2]). It should be noted that EOS of pure uranium is often used as an explicit constituent of combined EOS of hypo-stoichiometric uranium-oxygen mixture, the latter being described as a binary solution of U and $UO_2$ [10] (Fig.2 [12], Fig.5 [13], etc). It should be emphasized that serious uncertainty mentioned above on the presently known uranium EOS makes significant uncertainty of the whole hypo-stoichiometric part of total phase diagram for $UO_{2-x}$ ($0 < x < 2$). The same is true for similar problem in uranium-carbon, uranium-nitrogen and other uranium-bearing systems.

**3. Problem of theoretical estimation for critical point parameters. Uranium precedent**

The critical data (temperature and pressure) for uranium are too high for precise experimental study. At the same time *ab initio* theoretical approaches are ineffective because of the rather complicated electron structure of uranium [14] and also the still existing problem of adequate theoretical description of strong Coulomb interaction in non-ideal uranium plasma [1],[15]. The critical point parameters (CPP) for uranium are permanently estimated theoretically. Among theoretical approaches the dominating one is based on the assumption of strong correlation between CPP and low-temperature properties of condensed phase. There exist several versions of this approach which use either vaporization heat, or thermal expansion of liquid, or low-temperature vapor pressure as input quantity [3]. It should be stressed that whatever is used in frames of this approach does not matter - either any "primitive" form of high-temperature extrapolation of thermodynamic properties of "cold" condensed substance, such as Guldberg rule [3], or Kopp-Lang rule [16], or the low of "rectilinear diameter" [17], etc. (see [3],[4],[18]), or a variant of "principle corresponding states", or even any sophisticated forms of *modeling EOS* with free parameters. The point is which thermodynamic parameters of condense state to be preferably used in application of the principle corresponding states or in "calibration" of free parameters of modeling EOS.

To found this statement the special version of thermodynamic computer code, «SAHA-U» has been developed in present work as a new implementation of SAHA code-line [19],[1],[20]. The gas-liquid phase coexistence in uranium has been calculated by SAHA-U in frames of so-called the quasi-chemical representation, i.e. a microscopic description of vapor and liquid uranium as equilibrium partially ionized non-ideal plasma ("chemical picture") [21]. This approach proved to be successful for joint self-consistent description of non-congruent evaporation in uranium dioxide [22],[23],[2]. Two variants of present calculation by SAHA-U code correspond to two competing variants of calibration of its free parameters fitting either caloric or thermal properties of liquid uranium. In the first variant (notation "SAHA-U(*H*)") liquid density and handbook values of vaporization heat and Gibbs free energy (i.e. vapor pressure) of liquid uranium in melting point ($T = 1410$ K) were fitted. In the second variant (notation "SAHA-U(*T*)") density and thermal expansion coefficient [29],[30] of liquid uranium in melting point were fitted.

Among the approaches extrapolating low-temperature properties of condensed phase, the first and widely used version exploits the main *caloric quantity*, heat of vaporization as the basis for estimations of CPP. In accordance with the high value of uranium vaporization heat (~ 533 kJ/mol [24]), numerous attempts of such estimation predict relatively high values for the uranium critical temperature: $T_C$ ~ 11-13 $10^3$ K ([25],[26],[27],[4],[28] etc.). As it has been expected, the high value of critical temperature $T_c \approx 12800$ K is also obtained in calculations via SAHA-U code with caloric quantity used as the input calibration value (curve *4\** at Fig.1).

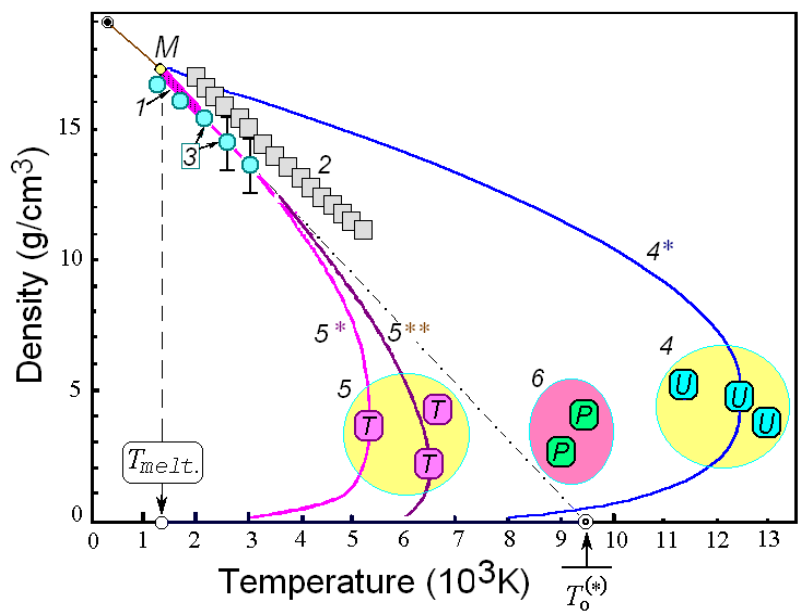

**Fig.1**. Density-temperature phase diagram of uranium.

<u>Experimental data on thermal EOS</u>: *1* – saturation curve $p \sim 0$ [30]; *2* – isobaric expansion ($p \cong 0.4$ GPa) [29] (after [34]); *3* – isobaric expansion [32];

<u>Theoretical predictions of critical point</u>: *4* – based on caloric EOS of condensed phase [25],[26],[27] [4] *etc.*; *5* – based on experimental data on thermal EOS of condensed phase [33],[5]; *6* – based on «plasma hypothesis» [36],[37].

<u>Reconstruction of coexistence curve</u>: *4\** – based on caloric EOS of condensed phase (code SAHA-U(*H*)); *5\** – based on thermal EOS [30] via principle of correspondent states and measured cesium coexistence curve [38],[39] as a reference system (after [5]); *5\*\** – based on thermal EOS [30] via procedure recommended in [40] (after [5]);

*Dash-dotted line* – linear extrapolation of liquid density, $\rho_{liquid}(T)$ [30] up to the limit $T_o^{(*)}$,

*M* – melting point; *NS* – density of solid uranium at $T = 273$ K.

The second version of this approach is based on parameters of *thermal EOS*, i.e. it extrapolates to high temperature experimentally measured density vs. temperature of liquid uranium, $\rho_{liquid}(T)$ [29],[30],[31],[32]. Good agreement should be emphasized for all four experiments in surprisingly high value of measured isobaric thermal expansion of liquid uranium (Fig.1). Correspondingly in contrast to the "caloric" way, the "thermal" way results in the significantly lower values of predicted critical temperature and noticeably indefinite values of critical pressure $T_C \sim$ 5-7 kK; $P_C \sim$ 0.01 - 0.5 GPa ([33],[34],[5]). The low value of critical temperature $T_c \approx 6840$ K is also obtained in calculations by SAHA-U code with *thermal* quantity being used as main input calibration value.

The third, an alternative approach, scarcely uses any empirical properties of condensed uranium in search of its critical point location. This approach is based on so-called "plasma hypothesis" of nature of critical state in metals [35], it postulates strong correlation of critical point parameters with *ionization potential(s)* of metal and it predicts for uranium $T_C \sim$ 9400 K [36],[37].

Obvious strong contradiction among three groups of theoretical predictions for the uranium critical temperature is illustrated in Fig.1. This discrepancy looks quite extraordinary in view of great applicative importance of uranium EOS.

## 4. Uncertainty in Uranium Saturated Pressure

Estimated from both thermal and caloric properties of liquid phase (Fig.1), the disaccord in uranium CPP becomes perfectly *incompatible* when considered jointly with Gibbs free energy of liquid. At low-temperature the latter is equivalent to the dependence of saturated vapor pressure on temperature, $p_s(T_s)$. For low temperature range this dependence is known experimentally and is recommended in thermodynamic handbooks (e.g., [41],[24]) as consolidation of total information on properties of condensed uranium. This recommendation for $T < 5000$ K is shown in Fig.2.

A standard approach of estimation for the critical pressure $p_C$ uses a linear extrapolation of this dependence in $\log p \leftrightarrow 1/T$ coordinates up to preliminary predicted value of critical temperature [3]. Numerous attempts of such estimations for uranium via "caloric way" (based on the vaporization heat) predict the relatively high values of critical pressure: $p_C \sim$ 0.5-1.0 GPa ([25],[26],[27],[4],[28] etc.). The large value of critical pressure $p_c \approx 845$ GPa has been also obtained in calculations by SAHA-U($H$) code with the caloric calibration.

In contrast to results mentioned above, the CPP estimation based on experimentally measured thermal expansion of liquid uranium [29],[30],[31],[32] could give high value of uranium critical pressure ($p_C \sim$ 0.4 GPa [33][34]). Almost the same value of uranium critical pressure ($p_C$ = 0.444 GPa) has been obtained in calculations by SAHA-U($T$) code with the thermal calibration (curve *7* at Fig.2) but with a saturation curve contradicting violently with the handbook data $p_s(T_s)$ [24]. On the other side, a superposition of low value of predicted uranium critical temperature ($T_C \sim$ 6 kK) based on thermal expansion with handbook $p_s(T_s)$ leads to the extra-low value of lower bound for critical pressure ($p_C \sim$ 0.01 GPa) [5],[6] and, consequently, to the extra-low value of critical compressibility factor, $Z_C \equiv (p/\rho RT)_C \sim$ 0.01 (!) as well as to the extra-low value of indicative density ratio $\varrho_C/\varrho_0 \sim$ 0.01 (!) ($\varrho_C$ and $\varrho_0$ – critical and normal density).

The third, alternative "plasma" approach [35], also predicts high value of uranium critical pressure ($p_C \sim$ 0.6 GPa [36]) but with a saturation curve [37] deviating significantly from the saturation curve of handbook [24]. A consolidated picture of all the discussed predictions of uranium critical pressure is exposed in Fig.2. One can conclude that the considerable uncertainty in predicted critical temperature of uranium (Fig.1) makes even more contrasting uncertainty in predictions of critical pressure presently known [5]. It should be noted that an attempt to combine the high value of predicted uranium critical pressure [33],[35],[37],[34] with the low-temperature experimental and handbook data on saturation curve [41],[24] leads to strong upward deviation of the high-temperature part of uranium saturation curve from the quasi-linear $\log p_s$-$1/T_s$ extrapolation of its low-temperature part (line 8 in Fig.2).

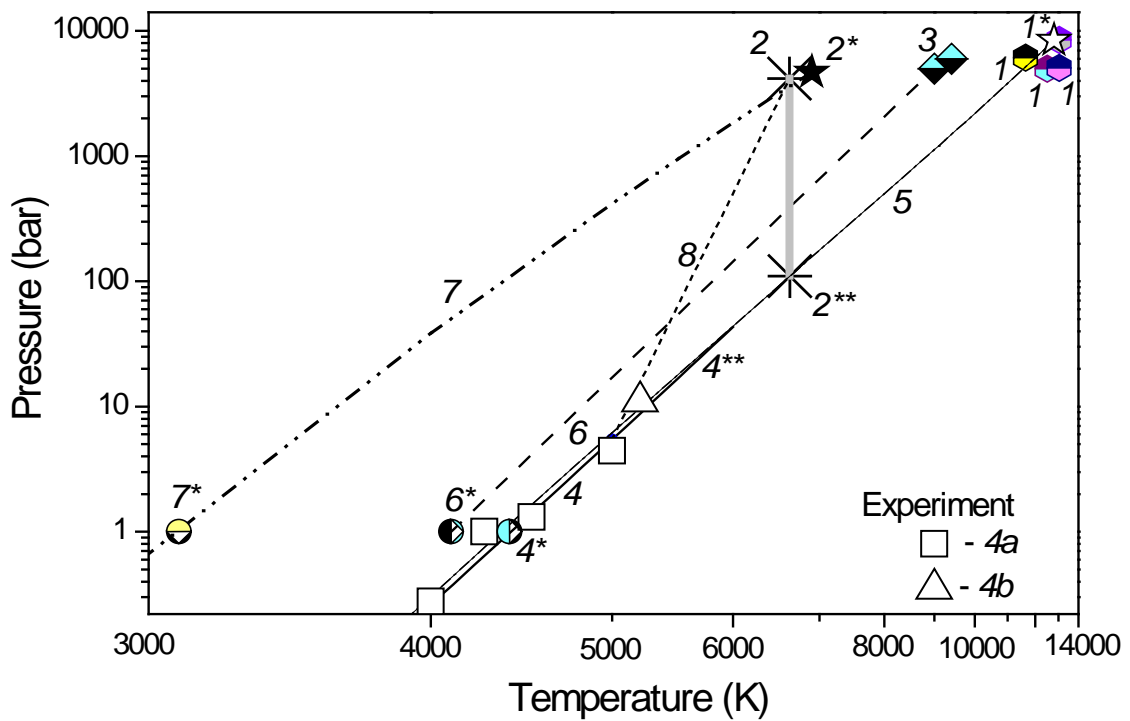

**Fig.2**. Pressure-temperature phase diagram of uranium.

<u>Theoretical predictions of critical point</u>: *1* – based on caloric EOS of liquid uranium [25],[27],[26],[4],[28]; *1\** – CP presently calculated via "caloric" variant of SAHA-U code; *2* – CP prediction [33],[34] based on experimental data [29] (thermal EOS); *2\** – present calculation via "thermal" variant of SAHA-U code; *2\*\** – low boundary of CP prediction [5][6] with low $T_C$ estimated via "thermal" way (region *5* at Fig.1), which is compatible with handbook vapor pressure [24]; *3* – CP predictions based on «plasma hypothesis» [36][37].

<u>Saturation curve</u>: *4* – handbook saturation curve [24] with boiling point (*4\**) and recommended extrapolation (*4\*\**); *4a* – experimental data [41]; *4b* – experimental boiling at $p$ = 1.1 MPa [31]; *5* – present calculation via "caloric" variant of SAHA-U code; *6* – saturation curve predicted in frames of «plasma hypothesis» [37], *6\** – boiling point recommended in [42]; *7* – present calculation via "thermal" variant of SAHA-U code; *7\** – corresponding boiling point; *8* – hypothetical part of vapor pressure curve connected handbook data (*4-4\**) with CP predicted in [33][34].

## 5. Uranium Critical Point Problem: is it exclusive or typical?

Significant discrepancy in uranium CPP predicted via the competing "caloric", "thermal" and "plasma" approaches (Fig.1-2) appears to be typical for group of metals [6]. There exists a ruling dimensionless parameter indicating definitely the possible discrepancy similar to that of uranium, it is the ratio $\eta$ of the indicative temperature of thermal EOS, $T_0^{(*)}$ (see Fig.1) in energy units, to the sublimation energy $\Delta_s H^0$.

$$\eta \equiv kT_0^{(*)}/\Delta_s H^0 \qquad (T_0^{(*)} \equiv T_{melt} + \alpha_P^{-1}|_{T=Tmelt}) \quad \{\alpha_P \equiv \varrho^{-1}(\partial\varrho/\partial T)_P\} \qquad (1)$$

A relatively low value of $\eta$ for uranium ($\eta_U \approx 0.15$) is typical for the whole group of "bad" metals: V, Co, Mo, W *etc.* [6] ($\eta \approx 0.16 - 0.19$). To compare the relatively high value of parameter $\eta$ (0.3 – 0.4) should be mentioned for "good" metal like Li, Cs, Al, Cu, etc. It should be also noted the remarkably low value of parameter $\eta$ for uranium dioxide ($\eta_{UO2} \approx 0.17$) that truly corresponds to well-known problem of high discrepancy between various theoretical estimations of critical point parameters of $UO_2$ [43],[2] etc.

## 6. In search for possible resolution of the uranium critical point problem

Summarizing up the discussion, one can conclude that this problem solving is required either (A) to disavow results of one (or more) basic experiments on thermodynamic properties of liquid uranium *or/and* (B) to assume at least one (or more) significant anomaly in properties of gas-liquid phase transition in uranium.

**(A)** Wrong experiments are the easiest explanation. In this case one has to assume: **(A*)** – there is a rough and simultaneous mistake in *four* experiments on measurements of liquid uranium density [29],[30],[31],[32], as well as in the measurements for uranium surface tension [30] (it is also indicates the extra-low critical temperature of U [5]) or/and **(A**)** – there is a mistake in experimental and handbook data for saturation pressure, $p_s(T_s)$ and evaporation heat, $\Delta H(T_s)$ [41],[24].

**(B)** – There is an anomalous (non-convex) form of density-temperature coexistence curve of uranium at high temperature (Fig.1), which could combine high value of experimentally measured thermal expansion of liquid uranium with high level of predicted critical temperature (*4* at Fig.1);

**(C)** – There is an anomalous large *upward non-linearity* of saturation curve, $p_s(T_s)$ in (log$p$–$1/T$) coordinates (Fig.2), which combines the low value of uranium vapor pressure at low temperature [41],[24], with the high level of predicted critical pressure [33],[35],[37],[34].

There are two variants within the validity of totally convex form of gas-liquid coexistence curve in $\varrho$–$T$ plane and validity of quasi-linear dependence in log$p_s$–$1/T_s$ coordinates:

**(D)** – There is an anomalous low value of the critical compressibility factor of uranium, $(pV/RT)_C \ll 1$;

**(E)** – There is an anomalous low value of the uranium ratio of critical to normal densities, $(\rho_C/\rho_0) \ll 1$.

In any case of either the wrong experiments (A) or the undetermined anomaly (B-E), some new decisive experiments are desirable for properties of liquid uranium and as well as for true parameters of its gas-liquid transition.

### 6.1. Possible solution and alternatives

**(A)** – <u>Is there an anomalous form of density-temperature diagram?</u>

In principle, it may be that the gas-liquid coexistence curve in density–temperature plane is *not totally convex* figure. It is so indeed, in particular, for the theoretically predicted high-temperature gas-liquid coexistence in uranium dioxide (Fig.3) [11] [23] (see [2] for details). It should be stressed that evaporation in $UO_{2\pm x}$ differs principally from that in "ordinary" substances because of non-congruency of this phase transition (coexistence of two phases with different stoichiometry). This is an intrinsic property of phase equilibrium in strongly interacting chemically reactive plasmas. The question is what physical reason could produce similar effect in metallic uranium. Could it be significant change in electronic structure of uranium ions and corresponding change in effective ion-ion interaction in liquid uranium during its expansion from melting to critical point?

**(B)** – <u>*Is there an anomalous high upward deviation of uranium saturation curve $p_s(T_s)$?*</u>

It is possible, in principle, that the saturation curve, $p_s(T_s)$, could strongly deviate upward from a generally accepted quasi-linear dependence in $\log p_s - 1/T_s$ coordinates. It is so indeed, in particular, for the predicted gas-liquid coexistence in uranium dioxide [11] [23] (see [2] for details). The physical reason for this remarkable deviation in $p_s(T_s)$ is again the non-congruence of gas-liquid coexistence in uranium dioxide. And the question still is what physical reason could produce similar effect in metallic uranium during its expansion from melting to critical point?

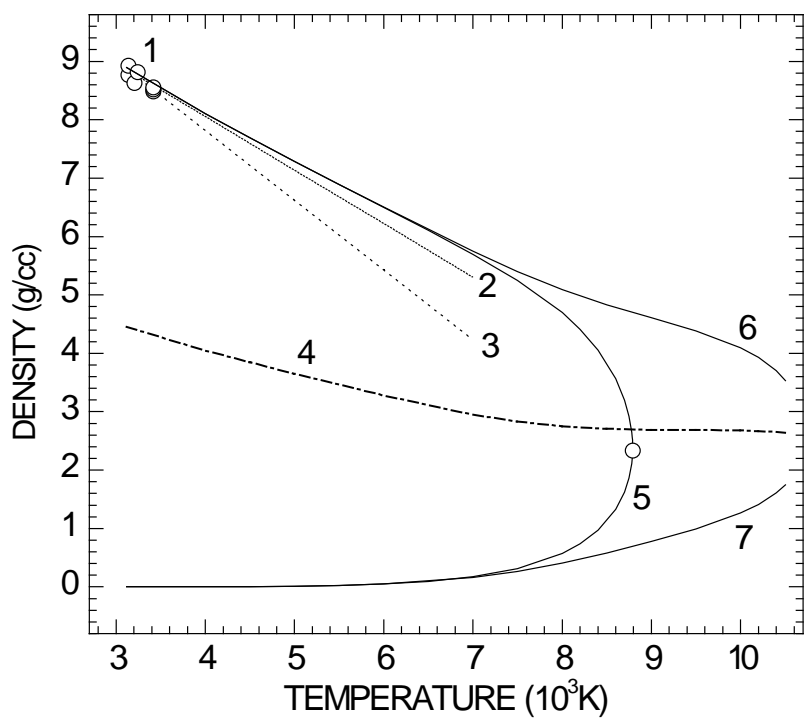

**Fig.3.** <u>Densities of coexisting gas and liquid phases for non-congruent boiling uranium dioxide</u>

*1* – experimental data on density of liquid $UO_{2.0}$ [47]; *2,3* – variants of extrapolation of *1*; *4* – the mean value of liquid and vapor densities (diameter); *5,6,7* - theoretical predictions within EOS [11]: *5* – standard form of coexisting curve (forced congruent equilibrium); *6,7* – coexisting liquid (*6*) and vapor (*7*) densities of non-congruent phase equilibrium in uranium dioxide within the Van der Waals approximation (see [2] for details).

**(C)** – *Is there an anomalous extra-low value of critical compressibility factor of uranium, $(pV/RT)_C \ll 1$ ?*

**(D)** – *Is there an anomalous extra-low value of the ratio of critical to normal densities of uranium, $(\rho_C/\rho_0) \ll 1$ ?*

It is possible, in principle, that the value of gas-liquid critical compressibility factor can be much lower even than that of alkali metals: $(pV/RT)_c \ll (pV/RT)_c^{(alkalis)} \approx 0,1 \div 0,2$. It is also possible that the ratio $(\rho_0/\rho_c)$ can be much higher than that of alkali metals: $\rho_0/\rho_c \gg (\rho_0/\rho_c)^{(alkalis)} \approx$ 5-10. This is the case, for example, in modified one-component plasma model, OCP(~) [44] in the formal limit of its ionic charge low value $z \ll 1$ [45] (see [46] for details).

## 7. Conclusions

Nowadays there is significant uncertainty in present knowledge on the uranium critical point parameters, the theoretical estimations being in striking contradiction. The version exploiting extrapolation of low temperature vapor pressure and heat of vaporization as the basis for estimations of CPP predicts high values for the uranium critical temperature. The version exploiting thermal expansion of liquid uranium results in significantly lower values of predicted critical temperature and noticeably indefinite values of critical pressure. This problem solution demands either to disavow results of one (or more) basic experiments on thermodynamic properties of liquid uranium or/and to assume at least one (or more) significant anomaly in properties of gas-liquid phase transition in uranium. The question is what physical reason could produce such anomalies in thermodynamic properties of metallic uranium during its expansion from melting to critical point? Solution of the problem requires both new theoretical efforts and new deciding experiments.

**Acknowledgements**

The work was supported by RAS scientific Programs "Physics and chemistry of extreme states of matter" and Grant № 775-04 of Atomic Energy Ministry of Russia. The work was partially supported also by Grants INTAS-93-66; CRDF № MO-011-0 and ISTC-2107.

**REFERENCES**

[1] V. Gryaznov, I. Iosilevskiy, Yu.G. Krasnikov, N.I. Kuznetsova, V.I. Kucherenko, G.V. Lappo, B.N. Lomakin, G.A. Pavlov, E.E. Son, V.E. Fortov, Thermophysics of Gas-Core Nuclear Reactor V. Ievlev, (Ed.), Atomizdat, Moscow, 1980 (in Russian).

[2] C. Ronchi, I.L. Iosilevski, E. Yakub, Equation of State of Uranium Dioxide, Springer, 2004, 366 pp.

[3] R.W. Ohse, H. Tippelskirch, High Temp.-High Pressures 9 (1977) 367.

[4] V.E. Fortov, A.N. Dremin, A.A. Leont'ev, High Temp. 13 (1975) 1072.

[5] Iosilevski I., in V. Fortov, (Ed.), Substance under Extreme Conditions, IVTAN, Moscow, 1991, p.106.

[6] Iosilevski I.L. in V. Fortov, (Ed.), Physics of matter under extreme conditions, Moscow, IPCP Chernogolovka, 2001, p.116 (in Russian).

[7] V.M. Ievlev, Bulletin of Russian Academy of Science (Izvestia RAS), № 6 (1977) 24.

[8] Rocket Engines and Energy Converters Based on Gas-Core Nuclear Reactor, A.S. Koroteev (Ed.), Mashinostroeniye, Moscow, 2002, 430 pp. (in Russian).

[9] A.E. Martin, R.K. Edwards, J. Phys. Chem., 69 (1965) 1788.

[10] J.F. Babelot, R.W. Ohse, M Hoch, J. Nucl. Mater. 137 (1986) 144.

[11] INTAS Project-93-66, JRC, ITU, Karlsruhe, Final Reports, Stage-I, 1997, Stage-II 1999.

[12] C. Gueneau, M. Baichi, D. Labroche, C. Chatillon, B. Sundman, J. Nucl. Mater. 304 (2002) 161.

[13] P. Chevalier, E. Fischer, B. Cheynet, J. Nucl. Mater. 303 (2002) 1.

[14] L.J. Radziemski, J.M. Mack, Physica 102 B (1980) 35.

[15] I.L. Iosilevskiy, Yu.G. Krasnikov, E.E. Son, V.E. Fortov, Thermodynamics and Transport in Non-Ideal Plasmas, MIPT Publishing, Moscow, 2002, 475 pp.; FIZMATLIT, Moscow, 2005, (in Russian).

[16] G. Lang, Z. Metallkd. 68 (1977) 213.

[17] L. Cailletet, E. Mathias, Compt. Rend. 104 (1887) 1563.

[18] H. Hess, H. Schneidenbach, Z. Metallkd. 87 (1996) 979.

[19] I. Iosilevski, V. Gryaznov, Report of Institute of Thermal Processes (ITP) N 27405, Moscow, 1972 (in Russian).

[20] V.K. Gryaznov, I.L. Iosilevskiy, V.E. Fortov, Nuclear Instruments & Methods in Physics Research A 415 (1998) 581.

[21] V.K. Gryaznov, I.L. Iosilevskiy, V.E. Fortov, in Shock Waves and Extreme States of Matter, V.E. Fortov, L.V. Altshuler, R.F. Trunin, A.I. Funtikov (Eds.), Springer, 2004, p. 437.

[22] V.K. Gryaznov, I.L. Iosilevski, A.M. Semenov, E.S. Yakub, V.E. Fortov, G.J. Hyland, C. Ronchi, Bulletin Russ. Acad. Sci. (Izvestia RAS), 63 (1999) 2258 (in Russian.)

[23] I. Iosilevski, G. Hyland, E. Yakub, C. Ronchi, Trans. Amer. Nucl. Soc. 81 (1999) 122; Int. J. Thermophys. 22 (2001) 1253; Contrib. Plasma Phys. 43 (2003) 316.

[24] Thermodynamic Properties of Individual Substances (Handbook), V.P. Glushko, L.V. Gurvich, G.A. Bergman, I.V. Veits, V.A. Medvedev, G.A. Khachkuruzov, V.S. Yungman (Eds.), v. IV, Nauka Publ., Moscow, 1982; Hemisphere Publ. New York, 1989; (see also the modern data collection in [11] [2])

[25] V. Grosse, J. Inorg. Nucl. Chem. 22 (1961) 23.

[26] D. Young, B. Alder, Phys. Rev. A 3 (1971) 364.

[27] G. Gathers, J. Shaner, D. Young, Phys. Rev. Lett. 33 (1974) 70.

[28] K. Hornung, J. Appl. Phys. 46 (1975) 2543.

[29] J. Shaner, Report UCRL-52352, LLNL, (1977); IV Int. High Press. Conference (1977).

[30] E.E. Shpil'rain, V.A. Fomin, V.V. Kachalov, High Temp. 26 (1988) 892.

[31] R.N. Mulford, Sheldon R.I., J. Nucl. Mater. 154 (1988) 268; ibid. 185 (1991) 297.

[32] M. Boiveneau, L. Arles, J. Vermenlen, Th. Threvenin, Physica B 190 (1993) 31.

[33] Young D. Soft Spheres Model for EOS, UCRL-52352, LLNL, Univ. California, 1977.

[34] G. Gathers, Rep. Progr. Phys. 9 (1986) 341.

[35] A. Likalter, Doklady of Russian Academy of Science 269 (1981) 96; High Temp. 23 (1985) 465.

[36] A. Likalter, Phys. Rev. B 53 (1996) 4386; Physica Scripta 55 (1997) 114.

[37] H. Hess, H. Schneidenbach, Z. Metallkd. 92 (2001) 399.

[38] S. Jungst, B. Knuth, F. Hensel, Phys. Rev. Lett. 55 (1985) 2160

[39] V.F. Kozhevnikov, Soviet JETF, 97 (1990) 541.

[40] L.P. Fillippov, High Temperature 22 (1984) 679; *ibid* 25 (1987) 1087.

[41] R. Hultgren, P.D. Desai, D.T. Hawkins, M. Gleiser, K.K. Kelley, D.D. Wagman, Selected Values of the Thermodynamic Properties of the Elements, Am. Soc. for Metals, Metals Park, OH (1973)

[42] T. Iida, R. Guthrie, Physical Properties of Liquid Metals, Clarendon Press, Oxford, 1993.

[43] R.W. Ohse, J-F. Babelot, C. Cercignani, J-P. Hiernaut, M. Hoch, , G.J. Hyland, J. Magill, J. Nucl. Mater. 130 (1985) 165.

[44] I. Iosilevski, High Temperature 23 (1985) 807.

[45] I. Iosilevski, A. Chigvintsev, in W. Kraeft and M. Schlanges (Eds.), Physics of Strongly Coupled Plasmas, World Scientific, Singapore/New Jersey/London, 1996, p.141.

[46] I.L. Iosilevskiy, Thermodynamics of non-ideal plasmas, in A. Starostin and I. Iosilevskiy (Eds.), Encyclopedia of Low-Temperature Plasma, v.VI (Supplement), Moscow, FIZMATLIT, 2005 (in Russian)

[47] W.D. Drotning, High Temp. – High Presure 13 (1981) 441.